# An Adjusted Likelihood Ratio Test for Separability in Unbalanced Multivariate Repeated Measures Data

Sean L. Simpson

*Department of Biostatistical Sciences, Medical Center Boulevard,*

*Winston-Salem, NC 27157-1063*

SLSimpso@wfubmc.edu

Sean L. Simpson is Assistant Professor, Department of Biostatistical Sciences, Wake Forest University School of Medicine, Winston-Salem, NC 27157 (E-mail: slsimpso@wfubmc.edu).

## ABSTRACT

We propose an adjusted likelihood ratio test of two-factor separability (Kronecker product structure) for unbalanced multivariate repeated measures data. Here we address the particular case where the within subject correlation is believed to decrease exponentially in both dimensions (e.g., temporal and spatial dimensions). However, the test can be easily generalized to factor specific matrices of any structure. A simulation study is conducted to assess the inference accuracy of the proposed test. Longitudinal medical imaging data concerning schizophrenia and caudate morphology illustrates the methodology.

## 1. INTRODUCTION

Multivariate repeated measures studies are characterized by data that have more than one set of correlated outcomes or repeated factors. Spatio-temporal data fall into this more general category since the outcome variables are repeated in both space and time. When analyzing multivariate repeated measures data, it is often advantageous to model the covariance separately for each repeated factor. This method of modeling the covariance utilizes the Kronecker product to combine the factor specific covariance structures into an overall covariance model. A covariance matrix is separable if and only if it can be written as $\Sigma = \Gamma \otimes \Omega$, where $\Gamma$ and $\Omega$ are factor specific covariance matrices (e.g. the covariance matrices for the temporal and spatial dimensions of spatio-temporal data respectively). This Kronecker product approach circumvents the need for a very large sample size required by the classical technique of taking $\Sigma$ to be unstructured. Another key advantage of the Kronecker product model lies in the ease of interpretation in terms of the independent contribution of every repeated factor to the overall within-subject error covariance matrix. It also has numerous computational advantages as detailed in Galecki [4], Naik and Rao [9], and Mitchell et al. [8].

Several tests have been developed to determine the appropriateness of a separable covariance model. Shitan and Brockwell [16] constructed an asymptotic chi-square test for separability. Likelihood ratio tests for separability were derived by Lu and Zimmerman [7], Mitchell et al. [8], Roy and Khattree [10-12], and Roy and Leiva [13]. Fuentes [3] developed a test for separability of a spatio-temporal process utilizing spectral methods. All of these tests were developed for balanced data. Also, with the exception of the tests proposed by Roy and Khattree [11,12] and Roy and Leiva [13], they were all developed for unstructured factor specific covariance matrices. Roy and Khattree [11] derived a test for the case where one factor specific matrix is compound symmetric and the other unstructured; while Roy and Khattree [12] developed a test for when one factor specific matrix has the discrete-time AR(1) structure and the other is unstructured. The test of Roy and Leiva, like those of Roy and Khattree [11,12], also provides a useful extension to the structured covariance case. Though, their test is also limited to the imposition of either a compound symmetric or discrete-time AR(1) structure on the factor specific matrices for balanced data. They note the importance of developing testing procedures for other correlation structures.

We propose an adjusted likelihood ratio test of Kronecker product covariance structure for unbalanced multivariate repeated measures data, i.e., $\Sigma_i = \sigma^2 \Gamma_i \otimes \Omega_i$ when an equal variance structure is assumed. Here we address the particular case where the within subject correlation is believed to decrease exponentially in both dimensions by assuming $\Gamma_i$ and $\Omega_i$ have the *linear exponent autoregressive* (LEAR) correlation structure presented in Simpson [17,18]. However, the test can be easily generalized to factor specific matrices of any structure. The LEAR model allows for an attenuation or acceleration of the exponential decay rate imposed by the continuous-time AR(1) structure, with the AR(1), compound symmetry, and MA(1) models being special cases of the LEAR structure. The inherent flexibility of the LEAR correlation structure allows it to accommodate a wide class of repeated measures data.

The testing procedure is presented and discussed in Section 2. A simulation study in Section 3 assesses the inference accuracy of the proposed test. Longitudinal medical imaging data concerning schizophrenia and caudate morphology illustrates the methodology in Section 4. We conclude with a summary discussion including planned future research in Section 5.

## 2. THE LIKELIHOOD RATIO TEST

### 2.1 Standard Test

We consider the following likelihood ratio test of separability for unbalanced data:

$$H_0: \boldsymbol{\Sigma}_i = \sigma^2 \boldsymbol{\Gamma}_i \otimes \boldsymbol{\Omega}_i;\ \boldsymbol{\Gamma}_i, \boldsymbol{\Omega}_i \,\text{LEAR} \quad \text{vs.} \quad H_1: \boldsymbol{\Sigma}_i \,\text{unstructured, positive definite.} \quad (1)$$

Suppose $\boldsymbol{y}_i$ is a $t_i s_i \times 1$ vector of $t_i s_i$ observations (e.g., $t_i$ temporal measurements and $s_i$ spatial measurements) on the $i^{th}$ subject $i \in \{1, \ldots, N\}$. We assume that $N > \max_i(t_i s_i)$. Let $\mathcal{C}(y_{ijl}, y_{ikl}) = \rho_{i\gamma;jk}$ and $\mathcal{C}(y_{ijl}, y_{ijm}) = \rho_{i\omega;lm}$ represent the temporal (or factor 1) and spatial (or factor 2) correlations respectively, where $\mathcal{C}(\,\cdot\,)$ is the correlation operator. Then for $\boldsymbol{\Gamma}_i = \{\rho_{i\gamma;jk}\}$ (the temporal/factor 1 correlation matrix) and $\boldsymbol{\Omega}_i = \{\rho_{i\omega;lm}\}$ (the spatial/factor 2 correlation matrix), the factor specific *linear exponent autoregressive* (LEAR) correlation structures are

$$\rho_{i\gamma;jk} = \mathcal{C}(y_{ijl}, y_{ikl}) = \begin{cases} \rho_\gamma^{d_{t;\min} + \delta_\gamma[(d(t_{ijl}, t_{ikl}) - d_{t;\min})/(d_{t;\max} - d_{t;\min})]} & j \neq k, \\ 1 & j = k \end{cases} \quad (2)$$

$$\rho_{i\omega;lm} = \mathcal{C}(y_{ijl}, y_{ijm}) = \begin{cases} \rho_\omega^{d_{s;\min} + \delta_\omega[(d(s_{ijl}, s_{ijm}) - d_{s;\min})/(d_{s;\max} - d_{s;\min})]} & l \neq m, \\ 1 & l = m \end{cases} \quad (3)$$

where $d(t_{ijl}, t_{ikl})$ and $d(s_{ijl}, s_{ikl})$ are the distances between measurement times and locations respectively, $(d_{t;\min}, d_{s;\min})$ and $(d_{t;\max}, d_{s;\max})$ are computational *constants* equal to the minimum and maximum number of temporal and spatial distance units across all subjects, $\rho_\gamma$ and $\rho_\omega$ are the correlations between observations separated by one unit of time and distance respectively, and $\delta_\gamma$ and $\delta_\omega$ are the decay speeds. Thus, the unknown parameters are $\boldsymbol{\tau} = \{\boldsymbol{\tau}_\gamma; \boldsymbol{\tau}_\omega\} = \{\delta_\gamma, \rho_\gamma; \delta_\omega, \rho_\omega\}$. We assume $0 \leq \rho_\gamma, \rho_\omega < 1$ and $0 \leq \delta_\gamma, \delta_\omega$. The $(d_{t;\min}, d_{s;\min})$ and $(d_{t;\max}, d_{s;\max})$ constants allow the model to adapt to the data and scale distance such that the multiplier of the decay speeds $\delta_\gamma$ and $\delta_\omega$, $(d(t_{ijl}, t_{ikl}) - d_{t;\min})/(d_{t;\max} - d_{t;\min})$ and $(d(s_{ijm}, s_{ijm}) - d_{s;\min})/(d_{s;\max} - d_{s;\min})$, is between 0 and 1 for computational purposes. One could also consider tuning these constants if necessary to address, for example, convergence issues. Simpson et al. [17,18] contains further details of the LEAR model.

Following the preceding notation, and assuming that $\boldsymbol{y}_i \sim N_{t_i s_i}(\boldsymbol{\mu}_i = \boldsymbol{X}_i\boldsymbol{\beta}, \sigma^2[\boldsymbol{\Gamma}_i(\boldsymbol{\tau}_\gamma) \otimes \boldsymbol{\Omega}_i(\boldsymbol{\tau}_\omega)])$ and is independent of $\boldsymbol{y}_{i'}$ for $i \neq i'$ where $\boldsymbol{\Gamma}_i$ and $\boldsymbol{\Omega}_i$ are defined in Equations 2 and 3, the likelihood and log likelihood functions under $H_0$ are given by

$$L(\boldsymbol{y}; \boldsymbol{\beta}, \sigma^2, \boldsymbol{\tau}) \tag{4}$$
$$= \prod_{i=1}^{N} (2\pi)^{-t_i s_i/2} |\sigma^2 \boldsymbol{\Gamma}_i \otimes \boldsymbol{\Omega}_i|^{-1/2} \exp\{-\boldsymbol{r}_i(\boldsymbol{\beta})'(\sigma^2 \boldsymbol{\Gamma}_i \otimes \boldsymbol{\Omega}_i)^{-1} \boldsymbol{r}_i(\boldsymbol{\beta})/2\}$$

and

$$l(\boldsymbol{y}; \boldsymbol{\beta}, \sigma^2, \boldsymbol{\tau}) \tag{5}$$
$$= -\frac{n}{2}\ln(2\pi) - \frac{1}{2}\sum_{i=1}^{N}\ln|\sigma^2\boldsymbol{\Gamma}_i \otimes \boldsymbol{\Omega}_i| - \frac{1}{2\sigma^2}\sum_{i=1}^{N}\boldsymbol{r}_i(\boldsymbol{\beta})'(\boldsymbol{\Gamma}_i \otimes \boldsymbol{\Omega}_i)^{-1}\boldsymbol{r}_i(\boldsymbol{\beta})$$
$$= -\frac{n}{2}\ln(2\pi) - \frac{1}{2}\sum_{i=1}^{N}\left(t_i s_i \ln(\sigma^2) + \ln|\boldsymbol{\Gamma}_i \otimes \boldsymbol{\Omega}_i|\right) - \frac{1}{2\sigma^2}\sum_{i=1}^{N}\boldsymbol{r}_i(\boldsymbol{\beta})'(\boldsymbol{\Gamma}_i \otimes \boldsymbol{\Omega}_i)^{-1}\boldsymbol{r}_i(\boldsymbol{\beta})$$

respectively, where $n = \sum_{i=1}^{N} t_i s_i$ and $\boldsymbol{r}_i(\boldsymbol{\beta}) = \boldsymbol{y}_i - \boldsymbol{X}_i\boldsymbol{\beta}$. The maximum likelihood (ML) estimates are derived following the approach used in Simpson et al. [18]. There they profile $\sigma^2$ out of the likelihood and utilize the profile log likelihood given by

$$l_p(\boldsymbol{y}; \boldsymbol{\beta}, \boldsymbol{\tau}) \tag{6}$$
$$= -\frac{1}{2}\sum_{i=1}^{N}\ln|\boldsymbol{\Gamma}_i \otimes \boldsymbol{\Omega}_i| - \frac{1}{2}n\ln\left[\sum_{i=1}^{N}\boldsymbol{r}_i(\boldsymbol{\beta})'(\boldsymbol{\Gamma}_i^{-1} \otimes \boldsymbol{\Omega}_i^{-1})\boldsymbol{r}_i(\boldsymbol{\beta})\right] - \frac{1}{2}n\ln\left(\frac{1}{n}\right) - \frac{n}{2}$$

To avoid computational issues it is best to use the equality

$$\ln|\boldsymbol{\Gamma}_i \otimes \boldsymbol{\Omega}_i| = s_i\ln|\boldsymbol{\Gamma}_i| + t_i\ln|\boldsymbol{\Omega}_i|$$

in case $|\boldsymbol{\Gamma}_i \otimes \boldsymbol{\Omega}_i|$ is close to zero.

The ML estimate of $\boldsymbol{\beta}$ can be expressed as

$$\widehat{\boldsymbol{\beta}}(\boldsymbol{\tau}) = \left(\sum_{i=1}^{M}\boldsymbol{X}_i'(\boldsymbol{\Gamma}_i^{-1} \otimes \boldsymbol{\Omega}_i^{-1})\boldsymbol{X}_i\right)^{-1}\left(\sum_{i=1}^{M}\boldsymbol{X}_i'(\boldsymbol{\Gamma}_i^{-1} \otimes \boldsymbol{\Omega}_i^{-1})\boldsymbol{y}_i\right); \tag{7}$$

however, there are no closed form expressions for the ML estimates of $\boldsymbol{\Gamma}_i$ and $\boldsymbol{\Omega}_i$ (which are functions of $\boldsymbol{\tau}$) and thus they are computed by utilizing the Newton-Raphson algorithm which requires the first and second partial derivatives of the profile log-likelihood in Equation 6. The derivations of the first partial derivatives are available from the author and more general forms of these derivatives can be found in Jennrich and Schluchter [5]. The second partial derivatives of the parameters are approximated by finite difference formulas (available in SAS [14]). The analytic second derivatives can also be derived explicitly. However, the approximations have proven very accurate. One could also employ finite difference formulas to approximate the first partial derivatives of the profile log-likelihood in Equation 6. Though, some accuracy may be lost in doing this.

After getting the estimates of $\boldsymbol{\Gamma}_i$, $\boldsymbol{\Omega}_i$, and $\boldsymbol{\beta}$ utilizing the Newton-Raphson algorithm, an estimate of $\sigma^2$ is calculated by substituting the estimates into

$$\widehat{\sigma}^2(\boldsymbol{\beta}, \boldsymbol{\tau}) = \frac{1}{n}\sum_{i=1}^{N}\boldsymbol{r}_i(\boldsymbol{\beta})'(\boldsymbol{\Gamma}_i^{-1} \otimes \boldsymbol{\Omega}_i^{-1})\boldsymbol{r}_i(\boldsymbol{\beta}), \tag{8}$$

which is the expression resulting from the initial profiling of $\sigma^2$ out of the likelihood.

The maximum of the likelihood under $H_0$ is then

$$
\max_{H_0} L(\boldsymbol{y}; \boldsymbol{\beta}, \sigma^2, \boldsymbol{\tau}) \tag{9}
$$
$$
= \exp\left\{-\frac{1}{2\widehat{\sigma}^2} \sum_{i=1}^{N} \boldsymbol{r}_i(\widehat{\boldsymbol{\beta}}_0)'(\widehat{\boldsymbol{\Gamma}}_i \otimes \widehat{\boldsymbol{\Omega}}_i)^{-1} \boldsymbol{r}_i(\widehat{\boldsymbol{\beta}}_0)\right\} \prod_{i=1}^{N} (2\pi)^{-t_i s_i/2} |\widehat{\sigma}^2 \widehat{\boldsymbol{\Gamma}}_i \otimes \widehat{\boldsymbol{\Omega}}_i|^{-1/2}
$$
$$
= \exp\left\{-\frac{1}{2\widehat{\sigma}^2} \sum_{i=1}^{N} \boldsymbol{r}_i(\widehat{\boldsymbol{\beta}}_0)'(\widehat{\boldsymbol{\Gamma}}_i^{-1} \otimes \widehat{\boldsymbol{\Omega}}_i^{-1}) \boldsymbol{r}_i(\widehat{\boldsymbol{\beta}}_0)\right\} \prod_{i=1}^{N} (2\pi\widehat{\sigma}^2)^{-t_i s_i/2} |\widehat{\boldsymbol{\Gamma}}_i|^{-s_i/2} |\widehat{\boldsymbol{\Omega}}_i|^{-t_i/2},
$$

where the estimates of $\boldsymbol{\Gamma}_i$, $\boldsymbol{\Omega}_i$, $\boldsymbol{\beta}_0$ (Equation 7), and $\sigma^2$ (Equation 8) are those resulting from the aforementioned maximum likelihood approach based on the profile log-likelihood in Equation 6. Given the imbalance in the data, the same algorithmic approach is also used to derive the maximum likelihood estimates of $\boldsymbol{\Sigma}_i$ and $\boldsymbol{\beta}_1$ under $H_1$. The ML estimate of $\boldsymbol{\beta}_1$ can be expressed as

$$
\widehat{\boldsymbol{\beta}}_1(\boldsymbol{\Sigma}_i) = \left(\sum_{i=1}^{M} \boldsymbol{X}_i' \boldsymbol{\Sigma}_i^{-1} \boldsymbol{X}_i\right)^{-1} \left(\sum_{i=1}^{M} \boldsymbol{X}_i' \boldsymbol{\Sigma}_i^{-1} \boldsymbol{y}_i\right), \tag{10}
$$

while as before there is no closed form expression for $\boldsymbol{\Sigma}_i$ and thus the Newton-Raphson algorithm is again employed to simultaneously solve for the estimates of $\boldsymbol{\Sigma}_i$ and $\boldsymbol{\beta}_1$ (Equation 10). The maximum of the likelihood under $H_1$ is then

$$
\max_{H_1} L(\boldsymbol{y}; \boldsymbol{\beta}, \boldsymbol{\Sigma}_i) \tag{11}
$$
$$
= \exp\left\{-\frac{1}{2} \sum_{i=1}^{N} \boldsymbol{r}_i(\widehat{\boldsymbol{\beta}}_1)' \widehat{\boldsymbol{\Sigma}}_i^{-1} \boldsymbol{r}_i(\widehat{\boldsymbol{\beta}}_1)\right\} \prod_{i=1}^{N} (2\pi)^{-t_i s_i/2} |\widehat{\boldsymbol{\Sigma}}_i|^{-1/2}.
$$

Consequently we have that the standard likelihood ratio is given by

$$
\Lambda = \frac{\max_{H_0} L}{\max_{H_1} L} = \frac{\exp\left\{-\frac{1}{2\widehat{\sigma}^2} \sum_{i=1}^{N} \boldsymbol{r}_i(\widehat{\boldsymbol{\beta}}_0)'(\widehat{\boldsymbol{\Gamma}}_i^{-1} \otimes \widehat{\boldsymbol{\Omega}}_i^{-1}) \boldsymbol{r}_i(\widehat{\boldsymbol{\beta}}_0)\right\} \prod_{i=1}^{N} \widehat{\sigma}^{-t_i s_i} |\widehat{\boldsymbol{\Gamma}}_i|^{-s_i/2} |\widehat{\boldsymbol{\Omega}}_i|^{-t_i/2}}{\exp\left\{-\frac{1}{2} \sum_{i=1}^{N} \boldsymbol{r}_i(\widehat{\boldsymbol{\beta}}_1)' \widehat{\boldsymbol{\Sigma}}_i^{-1} \boldsymbol{r}_i(\widehat{\boldsymbol{\beta}}_1)\right\} \prod_{i=1}^{N} (2\pi)^{-t_i s_i/2} |\widehat{\boldsymbol{\Sigma}}_i|^{-1/2}}. \tag{12}
$$

Under regularity conditions, $-2\ln\Lambda$ is asymptotically distributed as a $\chi_\nu^2$ random variable. The associated degrees of freedom $\nu$ is given by

$$\nu = \max_i \left( \frac{t_i s_i(t_i s_i + 1)}{2} \right) - 5 \tag{13}$$

since the $\max_i(t_i s_i(t_i s_i + 1)/2)$ gives the effective number of covariance parameters that has to be estimated under the alternative hypothesis. Thus, if the factor specific covariance matrices were assumed to be unstructured under the null, the associated degrees of freedom would be

$$\nu = \max_i \left( \frac{t_i s_i(t_i s_i + 1)}{2} \right) - \max_i \left( \frac{t_i(t_i + 1)}{2} + \frac{s_i(s_i + 1)}{2} - 1 \right). \tag{14}$$

It is important to note that if any of the covariance parameters, $\{\sigma^2; \delta_\gamma, \rho_\gamma; \delta_\omega, \rho_\omega\}$ in our case, reside on the boundary of their parameter space then the asymptotic distribution of $-2\ln\Lambda$ becomes a mixture of $\chi^2$ distributions as discussed in Self and Liang [15].

For small samples, or when $N$ is not much greater than $\max_i(t_i s_i)$, using the empirical distribution of $-2\ln\Lambda$ as discussed in Lu and Zimmerman [7] has appeal (though, it is important to note that they only worked with balanced data). However, determining the empirical distribution may not always be computationally feasible. Modifying the critical value of the associated $\chi^2$ distribution serves as another approach. Mitchell et al. [8] provide one such critical value adjustment.

## 2.2 Adjusted Tests

Here we introduce two critical value adjustments for the LRT to deal with the case when $N$ is not much greater than $\max_i(t_i s_i)$. The first is adapted from the adjustment discussed in Mitchell et al. [8] which is based on the ratio of the mean of the LRT with its asymptotic mean. Their adjustment (for balanced data with unstructured covariance matrices) is defined as

$$k = \frac{-N\big(ts \ln 2 + \sum_{j=1}^{ts} \psi(0.5(N-j)) - ts \ln N\big) - (N/(N-1))(t(t+1)/2 + s(s+1)/2 + ts - 1)}{ts(ts+1)/2 - t(t+1)/2 - s(s+1)/2 + 1},$$

(15)

where $\psi$ is the digamma function (Kocherlakota et al. [6] has details). Then $-2\ln\Lambda \approx k\chi_\nu^2$. We define an analog of this adjustment for the test statistic in Equation 12 as

$$k_1 = \frac{-N\left(\max_i(t_i s_i) \ln 2 + \sum_{j=1}^{\max_i(t_i s_i)} \psi(0.5(N-j)) - \max_i(t_i s_i) \ln N\right) - (N/(N-1))\left(\max_i(t_i s_i) + 5\right)}{\max_i\left(\dfrac{t_i s_i(t_i s_i + 1)}{2}\right) - 5}.$$

(16)

We also define another, less conservative and more straightforward, adjustment as

$$k_2 = N / \left(N - \max_i(t_i s_i)\right).$$

(17)

This modification has the nice property that $k_2 \to 1$ for $N >> \max_i(t_i s_i)$ (i.e., it converges to the standard, unadjusted test). This is not the case for $k_1$ which fluctuates dramatically, and can even become negative (in extreme cases), for various $\left\{N, \max_i(t_i s_i)\right\}$ combinations.

## 3. SIMULATION STUDY

To assess the empirical performance of the likelihood ratio test based on the $k_1$ and $k_2$ adjustments, unbalanced multivariate repeated measures data were generated assuming the Kronecker product LEAR correlation structure with $\boldsymbol{\rho} = [\,\rho_\gamma \quad \rho_\omega\,]' = [\,0.8 \quad 0.8\,]'$, $\boldsymbol{\delta} = [\,\delta_\gamma \quad \delta_\omega\,]' = [\,(d_{t;\max} - d_{t;\min})/4 \quad (d_{s;\max} - d_{s;\min})/4\,]'$, and $\sigma^2 = 1$. Simulated test size at the $\alpha = 0.05$ level was examined for the test given in Equation 1 with sample sizes of $N = 80, 120, 160,$ and $200$. We set $\max_i(t_i) = 3, 5,$ or $7$, and $\max_i(s_i) = 4$ in order to facilitate convergence with an unstructured covariance model fit, with $(t_i \cdot s_i)\,\epsilon\,[1, 28]$ observations each at two-unit distance intervals. Two mean model scenarios were considered corresponding to signal-to-noise ratios (SNR) of none and moderate/high respectively: 1) $\beta = 0$ (one group with mean 0) and 2) $\boldsymbol{\beta} = [3.5, 3.5]'$ (one reference group with one additional group). The $R^2$ statistic presented in Edwards et al. [2], denoted $R_\beta^2$, was used as a proxy for SNR. Their statistic measures the multivariate association between the repeated outcomes and mean model in a repeated measures setting. Scenario 1 has $R_\beta^2 = 0$ (no signal, only noise), and Scenario 2 has an average $R_\beta^2$ value of 0.72 (moderate/high SNR) across all parameter combinations. Each simulation for the varying sample sizes and number of observations consisted of 5,000 realizations.

Table 1 shows the results of the simulations for $N \in \{80, 120, 160, 200\}$ and $\max_i(t_i) \in \{3, 5, 7\}$. The table contains simulated test size (target $\alpha = 0.05$) for the unadjusted and $k_2$-adjusted likelihood ratio test of $H_0$: $\boldsymbol{\Sigma}_i = \sigma^2 \boldsymbol{\Gamma}_i \otimes \boldsymbol{\Omega}_i$; $\boldsymbol{\Gamma}_i, \boldsymbol{\Omega}_i$ LEAR vs. $H_1$: $\boldsymbol{\Sigma}_i$ unstructured, positive definite. The $k_1$ adjustment was overly conservative (test size = 0.000 for all conditions) and thus left out of the results. There was also a lack of convergence under the alternative for $N = 80$ and $\max_i(t_i) \in \{5, 7\}$, thus there are no results to report for these conditions. This lack of convergence likely stems from the fact that $N = 80$ is not a large enough sample size to support the estimation of an unstructured covariance matrix with up to 210 and 406 parameters for $\max_i(t_i) = 5$ and 7 respectively. For an analyst, this should be a sign that the data does not support the given model, and a more parsimonious covariance model, like a Kronecker product structure, should be employed.

As evidenced by the results for $N = 80$ and 120, the sample size needs to be much larger than the maximum number of observations in order for test size to be controlled with the unadjusted test. The $k_2$-adjusted test controls test size across all parameter combinations except when $N = 120$ and $\max_i(t_i) = 7$, though it still far outperforms the standard test in this case. It is important to note that for this $N = 120$ and $\max_i(t_i) = 7$ case, the convergence rate was only moderate (under the alternative), thus these results should be viewed with caution. As previously mentioned, a more parsimonious covariance model should generally be fitted when these convergence issues arise. This special case $\left( N = 120 \text{ and } \max_i(t_i) = 7 \right)$ notwithstanding, the $k_2$-adjusted test also maintains a relatively consistent test size across all parameter combinations while the test size of the unadjusted test varies widely. The signal-to-noise ratio appears to have no influence on test size control for either test, paralleling the findings in Roy and Khattree [11].

## 4. EXAMPLE: SCHIZOPHRENIA AND CAUDATE MORPHOLOGY

The data include longitudinal MRI scans of the left caudate for 240 schizophrenia patients and 56 controls. The surface of each object extracted from the images was parameterized via the m-rep method as described in Styner and Gerig [19]. The caudate shape was determined as a 3 x 7 grid of mesh points (see Figure 1). Data were reduced to one outcome measure: *Radius* in cm as a measure of local object width (21 locations per caudate, $\max_i(s_i) = s = 21$). The distance between two radii for a given subject was calculated as the mean Euclidian distance over all images. Scans were taken up to 47 months post-baseline with the median and maximum number of scans per subject being 3 and 7 respectively, thus $\max_i(t_i) = 7$. As evidenced by information criteria and observed vs. predicted correlation plots, Simpson et al. [18] showed that the Kronecker product LEAR model provides a good fit to these data. However, no formal test of separability was performed as none was available that could accommodate this covariance structure or the imbalance in temporal measurements.

Here we apply our $k_2$-adjusted likelihood ratio test of separability to these data in order to determine whether the Kronecker product LEAR structure is appropriate. Due to convergence issues with an unstructured covariance fit, we had to pick four (out of the

21) representative spatial locations. We assume a significance level of $\alpha = 0.05$. The test statistic, $-2\ln\Lambda$, is equal to 699.34 (degrees of freedom = 401, $k_2 = 1.104$), corresponding to a p-value $< 0.0001$. Thus, we reject the null hypothesis and conclude that the Kronecker product LEAR structure does not provide a better fit than an unstructured matrix. The apparent inadequacy of the Kronecker fit could be due to a number of reasons. As mentioned by Cressie and Huang [1], patterns of interaction among the various factors cannot be modeled when utilizing a Kronecker product structure. Thus, it could be the case that the inadequate fit results from the temporal variability varying by spatial location or conversely the spatial variability varying by time. It might also be the case that with the full set of spatial observations the Kronecker model would prove the better fit. However, the large number of observations necessitates data reduction in order to fit an unstructured covariance. Herein lies one major advantage of the Kronecker model, namely it's ability to accommodate large numbers of observations that a single unstructured model cannot. In order to gain a better understanding of the disparate fits, we examine the following estimates (each multiplied by 100) for subject $i = 4$ ($t_4 = 3, s = 4$):

$$\hat{\sigma}^2\hat{\mathbf{\Gamma}}_4\otimes\hat{\mathbf{\Omega}} = \begin{pmatrix} 1.36 & 0.49 & 0.42 & 0.24 & 1.08 & 0.39 & 0.34 & 0.19 & 1.07 & 0.38 & 0.33 & 0.19 \\ 0.49 & 1.36 & 0.28 & 0.33 & 0.39 & 1.08 & 0.22 & 0.26 & 0.38 & 1.07 & 0.22 & 0.26 \\ 0.42 & 0.28 & 1.36 & 0.18 & 0.34 & 0.22 & 1.08 & 0.14 & 0.33 & 0.22 & 1.07 & 0.14 \\ 0.24 & 0.33 & 0.18 & 1.36 & 0.19 & 0.26 & 0.14 & 1.08 & 0.19 & 0.26 & 0.14 & 1.07 \\ 1.08 & 0.39 & 0.34 & 0.19 & 1.36 & 0.49 & 0.42 & 0.24 & 1.08 & 0.39 & 0.34 & 0.19 \\ 0.39 & 1.08 & 0.22 & 0.26 & 0.49 & 1.36 & 0.28 & 0.33 & 0.39 & 1.08 & 0.22 & 0.26 \\ 0.34 & 0.22 & 1.08 & 0.14 & 0.42 & 0.28 & 1.36 & 0.18 & 0.34 & 0.22 & 1.08 & 0.14 \\ 0.19 & 0.26 & 0.14 & 1.08 & 0.24 & 0.33 & 0.18 & 1.36 & 0.19 & 0.26 & 0.14 & 1.08 \\ 1.07 & 0.38 & 0.33 & 0.19 & 1.08 & 0.39 & 0.34 & 0.19 & 1.36 & 0.49 & 0.42 & 0.24 \\ 0.38 & 1.07 & 0.22 & 0.26 & 0.39 & 1.08 & 0.22 & 0.26 & 0.49 & 1.36 & 0.28 & 0.33 \\ 0.33 & 0.22 & 1.07 & 0.14 & 0.34 & 0.22 & 1.08 & 0.14 & 0.42 & 0.28 & 1.36 & 0.18 \\ 0.19 & 0.26 & 0.14 & 1.07 & 0.19 & 0.26 & 0.14 & 1.08 & 0.24 & 0.33 & 0.18 & 1.36 \end{pmatrix}$$

$$
\widehat{\boldsymbol{\Sigma}}_4 = \left(\begin{array}{cccc|cccc|cccc}
1.40 & 0.54 & 0.52 & 0.31 & 1.09 & 0.44 & 0.55 & 0.19 & 1.05 & 0.35 & 0.65 & 0.23 \\
0.54 & 1.37 & 0.61 & 0.70 & 0.27 & 0.98 & 0.56 & 0.48 & 0.37 & 1.11 & 0.60 & 0.47 \\
0.52 & 0.61 & 2.00 & 0.34 & 0.65 & 0.55 & 1.80 & 0.21 & 0.77 & 0.61 & 1.88 & 0.27 \\
0.31 & 0.70 & 0.34 & 1.04 & 0.24 & 0.63 & 0.23 & 0.63 & 0.21 & 0.64 & 0.30 & 0.70 \\
\hline
1.09 & 0.27 & 0.65 & 0.24 & 1.63 & 0.52 & 0.75 & 0.21 & 1.22 & 0.28 & 0.81 & 0.19 \\
0.44 & 0.98 & 0.55 & 0.63 & 0.52 & 1.19 & 0.56 & 0.56 & 0.42 & 1.01 & 0.60 & 0.52 \\
0.55 & 0.56 & 1.80 & 0.23 & 0.75 & 0.56 & 2.14 & 0.16 & 0.82 & 0.61 & 1.93 & 0.20 \\
0.19 & 0.48 & 0.21 & 0.63 & 0.21 & 0.56 & 0.16 & 0.75 & 0.16 & 0.50 & 0.18 & 0.61 \\
\hline
1.05 & 0.37 & 0.77 & 0.21 & 1.22 & 0.42 & 0.82 & 0.16 & 1.48 & 0.47 & 0.91 & 0.22 \\
0.35 & 1.11 & 0.61 & 0.64 & 0.28 & 1.01 & 0.61 & 0.50 & 0.47 & 1.30 & 0.61 & 0.53 \\
0.65 & 0.60 & 1.88 & 0.30 & 0.81 & 0.60 & 1.93 & 0.18 & 0.91 & 0.61 & 2.19 & 0.24 \\
0.23 & 0.47 & 0.27 & 0.70 & 0.19 & 0.52 & 0.20 & 0.61 & 0.22 & 0.53 & 0.24 & 0.90
\end{array}\right).
$$

As evidenced by the changing spatial covariance pattern (among the 4 caudate radii) across the 3 time points (the 3 $4 \times 4$ blocks along the diagonal of $\widehat{\boldsymbol{\Sigma}}_4$) under the alternative, there is most likely a space $\times$ time interaction which cannot be modeled by a separable structure. In other words, the spatial covariance pattern is different at different time points. Though, overall, the separable LEAR model seems to provide a reasonable approximation to the completely unstructured model with $78 - 5 = 73$ fewer parameters for this subject. Thus, its use for the full data $\left(\text{where } \max_i(t_i s_i) = 147\right)$ seems acceptable, especially given the inability of the alternative to handle the data's dimensionality.

## 5. DISCUSSION

We have presented an adjusted likelihood ratio test for separability that accommodates unbalanced multivariate repeated measures data. More specifically, we derived the test for the case where the within-subject correlation is believed to decrease exponentially for both factors by assuming a Kronecker product LEAR correlation model. Due to its parsimonious structure, the model is especially attractive for the high dimension, low sample size cases that are so common in medical imaging and various kinds of "-omics" data. Thus, development of hypotheses concerning its appropriateness are paramount. As evidenced by the simulation results, the adjusted test outperforms the standard test when $N$ is not much greater than $\max_i(t_i s_i)$ and converges to the standard test for $N >>$ $\max_i(t_i s_i)$. Future research examining alternative test modifications to improve test size control, including adjustments to the degrees of freedom, will prove very useful. As mentioned by Roy and Leiva [13], developing similar testing procedures for other correlation structures is also important.

## ACKNOWLEDGEMENTS

The author thanks Lloyd Edwards from the Department of Biostatistics at the University of North Carolina at Chapel Hill for his insight and suggested edits that greatly improved the presentation of material. The author also thanks the editor, associate editor, and referee for their comments that considerably improved the paper.

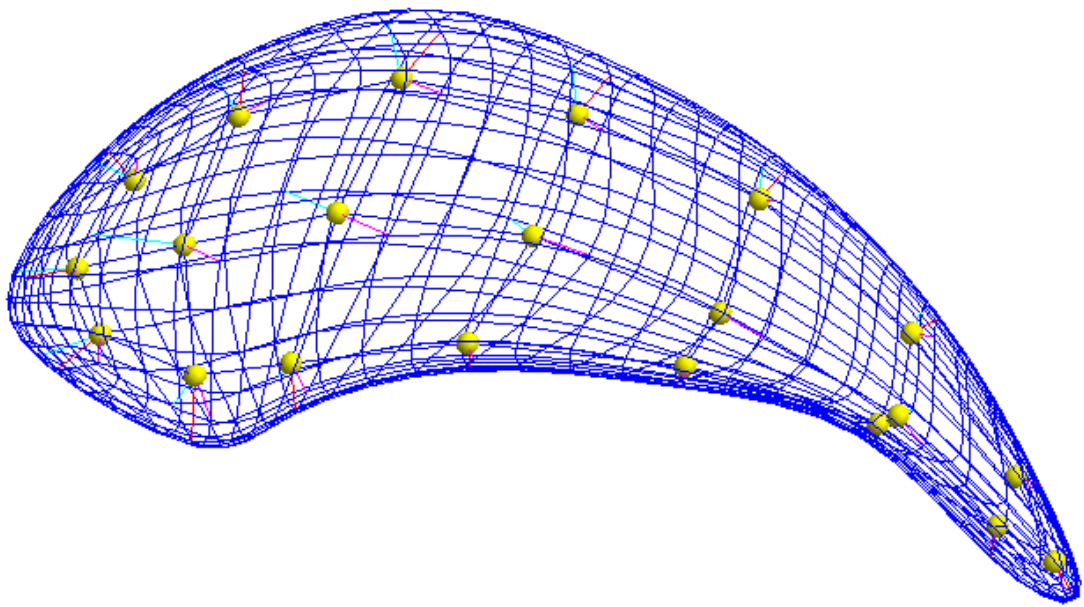

Figure 1. M-rep Shape Representation Model of the Caudate.

Table 1.  Simulated Test Size for Target $\alpha = 0.05$
5,000 realizations, $\max_{i}(s_i) = 4$

| SNR | $N$ | $\max(t_i)$ | df | Test Size[a] | |
|---|---|---|---|---|---|
| | | | | Adj. LRT | LRT |
| None | 80 | 3 | 73 | 0.041 | 0.170 |
| | | 5 | 205 | *[b] | *[b] |
| | | 7 | 401 | *[b] | *[b] |
| | 120 | 3 | 73 | 0.003 | 0.018 |
| | | 5 | 205 | 0.003 | 0.098 |
| | | 7 | 401 | 0.287[c] | 0.849[c] |
| | 160 | 3 | 73 | 0.000 | 0.002 |
| | | 5 | 205 | 0.000 | 0.002 |
| | | 7 | 401 | 0.000 | 0.031 |
| | 200 | 3 | 73 | 0.000 | 0.000 |
| | | 5 | 205 | 0.000 | 0.000 |
| | | 7 | 401 | 0.000 | 0.000 |
| Moderate/High | 80 | 3 | 73 | 0.040 | 0.161 |
| | | 5 | 205 | *[b] | *[b] |
| | | 7 | 401 | *[b] | *[b] |
| | 120 | 3 | 73 | 0.005 | 0.020 |
| | | 5 | 205 | 0.003 | 0.095 |
| | | 7 | 401 | 0.294[c] | 0.844[c] |
| | 160 | 3 | 73 | 0.000 | 0.002 |
| | | 5 | 205 | 0.000 | 0.002 |
| | | 7 | 401 | 0.000 | 0.035 |
| | 200 | 3 | 73 | 0.000 | 0.000 |
| | | 5 | 205 | 0.000 | 0.000 |
| | | 7 | 401 | 0.000 | 0.000 |

[a]SE $< 0.0064$
[b]Did not converge
[c]Moderate convergence rate